\documentclass[oldversion]{aa}  
\usepackage{graphicx}
\usepackage{amsmath,amssymb}
\usepackage{epsfig}                            %%temporary -> graphicx's includegraphics*
\usepackage{natbib}
\usepackage{txfonts}
\begin{document}
    \title{Doppler imaging an X-ray flare on the ultrafast rotator BO Mic
          \thanks{Based on observations obtained at the ESO VLT Obs. No. 078.D-0865(A) 
                  and {\it XMM-Newton} Obs. Id. 0400460301 and  0400460401}
                        %the European Southern Observatory using the VLT}
         }

  \subtitle{A contemporaneous multiwavelength study using XMM-Newton and VLT}

   \author{U. Wolter
          \inst{1},
           J. Robrade \inst{1}, 
          J.H.M.M. Schmitt \inst{1}
          %% \fnmsep\thanks{Just to show the usage of the elements in the author field}
          \and
          J.U. Ness \inst{2}
          }
   \authorrunning{Wolter et al.}   

   %\offprints{U. Wolter, \\ \email{uwolter@hs.uni-hamburg.de}}

   \institute{Hamburger Sternwarte, Gojenbergsweg 112, D-21029 Hamburg, Germany \\
              \email{uwolter@hs.uni-hamburg.de, jrobrade@hs.uni-hamburg.de, jschmitt@hs.uni-hamburg.de}
       \and
              Arizona State Universtity, PO Box 871404, Tempe, AZ, 85287-1404, USA
              \email{Jan-Uwe.Ness@asu.edu}
             }
   \date{Received / Accepted}  

  \abstract   %no sections: ``oldstyle'' above
   {We present an analysis of contemporaneous photospheric, chromospheric and coronal structures on the
    highly active K-dwarf star BO Mic (Speedy Mic). 
    %We localize a moderate flare in the stellar atmosphere and study its energetics, 
    %size     and thermal behaviour. 
   We concentrate on a moderate flare that we localize in the stellar 
   atmosphere and study its energetics, size and thermal behavior.
   The analysis is based on strictly 
   simultaneous X-ray, UV- and optical observations carried out by XMM-Newton and the VLT.
   We use Doppler imaging and related methods
   for the localization of features.
   Based on X-ray spectroscopy we study the 
   the coronal plasma in and outside the flare.
   The flare emits in X-rays and UV, but is not detected in white light;
    it is located at intermediate latitude between an extended spot group and 
    the weakly spotted pole. We estimate its height to be below 0.4~stellar radii,
    making it
    clearly distinct in longitude and height from the prominences found 
    %at heights of about three stellar radii.  
    about two stellar radii above the surface.  
    %While BO~Mic's photospheric brightness is modulated due to extended
    %starspots,
    In contrast to BO Mic's photospheric brightness,
    neither its chromospheric nor its X-ray emission show a pronounced
    rotational modulation.}
    %--ok-- shorten abstract

  % {ok importance of continous and strictly contemporaneous observations}
  %  {}


   \maketitle
%
%________________________________________________________________

%________________________________________________________________

\section{Introduction}
\vspace*{-0.75\medskipamount}
The level of stellar activity is related to the efficiency of a star's dynamo,
operated by an interplay of convection and rotation, with faster rotation resulting in higher activity for 
a given type of star (\citealt{Schrijver00}).  Extreme activity phenomena are observed on
the ultrafast rotating, highly active, young K-dwarf star 
%apparently single 
BO~Mic (K2V, \mbox{$P_{\mathsf{rot}}=0.380\pm0.004$~days}, 
\mbox{$v\,\sin{i} \approx 135$~km/s},
$d=44.5$~pc), 
nicknamed \mbox{``Speedy Mic''}.
BO~Mic produced the largest X-ray flare observed during the \textsc{Rosat} all-sky survey, 
with a peak flux of  \mbox{$\approx 9\cdot10^{31}$~erg/s}
in the \textsc{Rosat} PSPC passband (\mbox{$0.1-2.4$~keV}, \citealt{Kurster95}).
Follow-up observations of BO~Mic showed its very high projected rotational velocity 
%(\mbox{$v\,\sin{i} \approx 135$~km/s}) 
and short rotation period.
%, allowing highly resolved reconstructions of its surface 
%from observations covering only short time spans. 
Subsequent Doppler imaging studies revealed a richly spotted photosphere
and reconfigurations of active regions during only a few stellar 
rotations (\citealt{Wolter05} = WSW~2005, \citealt{Barnes05}), 
making \textit{strictly} simultaneous and coherent observations 
%covering short times 
mandatory. 
We observed BO~Mic during two consecutive nights
on 2006 October 13/14 and 14/15 simultaneously with {\it XMM-Newton}
and \textsc{Vlt/Uves} at ESO Paranal. The observations were scheduled to maximize 
simultaneous and continuous visibility of the target by both instruments.
%In this letter we report first results of 
%our campaign, presenting a coherent 'snapshot' of BO~Mic's active regions
%that includes the localization and analysis of a moderate flare.
%--ok fit in changes 

In this letter we report the first results of our campaign, presenting a
coherent ``snapshot'' of BO~Mic's active regions that concentrates
on the localization and analysis of a moderate flare.
% prominences ref Dunstone 2005, 2006

%ok fill 1st page space ?

%\vspace*{-1.3\medskipamount}
\section{Observations and data analysis}
\label{sec:Obs+Analysis}
\vspace*{-0.75\medskipamount}

%% ad phases:
%% print, [0.485, 0.695, 1.478, 1.693]/0.380  --> 1.27632  1.82895  3.88947  4.45526
Our \textsc{Vlt/Uves} observations lasted two half nights,
%JD$_0+0.485$ to JD$_0+0.695$ and JD$_0+1.478$ to JD$_0+1.693$,
JD$_0+0.49$ to $0.70$ and JD$_0+1.48$ to $1.69$
with JD$_0 = 2454022.0$.
% used as rotation phase zero point in the following.
The {\it XMM-Newton} observations exceeded these time spans, lasting from
from JD$_0+0.48$ to $0.79$ and from  JD$_0+1.47$ to $1.73$. In total
% 0.479 to 0.789
%JD~$2454023.467$ to JD~$2454023.727$, 
%while \textsc{Uves} observations covered JD~$2454022.485$ to JD~$2454022.695$ and 
%JD~$2454023.478$ to JD~$2454023.693$.
we obtained about 50~ks of X-ray data and 32 exposures 
%of roughly 1\,ks each 
of optical photometry from the optical monitor (OM) onboard {\it XMM-Newton}.
Our 142 \textsc{Uves}~spectra completely cover two semi-rotations of BO~Mic with some phase overlap.
%between them. 
  %start and end of the observations.
  %ultrafast readout
%, i.e. two half nights from dusk until

\textsc{Uves} was operated in dichroic mode, covering the spectral range 
3260 to 9460 \AA\ 
%with a gap between 4450 and~5650~\AA\ 
with a spectral resolution of $\lambda / \Delta\lambda\approx40\,000$.
%(DIC2, 390 and 760nm).
\textsc{Uves} exposures lasted between 200 and 250~s, 
separated by the CCD readout time of 10~s.
The spectra were reduced using the package \textsc{Reduce} 
described in \citet{Piskunov02}.
We carried out an approximate flux calibration using synthetic spectra from \citet{Haus99}. 

All {\it XMM-Newton} instruments obtained useful data of BO~Mic; here we only present data
gathered by the EPIC (European Photon Imaging Camera) and the OM (optical monitor).
The OM was operated with the visual grism in the wavelength range 3000-7000~\AA. 
Due to the brightness of BO~Mic, the OM~grism spectra are affected by 
coincidence loss, and we could not make use of either the full spectral resolution
or of the absolute flux calibration, thus we had to restrict the analysis to relative spectrophotometry.
All {\it XMM-Newton} data were reduced using the Science Analysis
System (SAS) version~7.0 (\citealt{sas}) with standard selection and filtering criteria.
X-ray light curves are background subtracted, based on nearby source-free regions.
Spectral analysis of the EPIC data was carried out with XSPEC version 11.3 (\citealt{xspec96}); for
spectral fitting purposes we used a multi-temperature model, assuming the emission spectrum of
a collisionally ionized, optically thin gas as calculated with the APEC code (\citealt{apec})
with abundances modeled once for the complete data set.  
%Abundances are modelled for the complete data and kept fixed to investigate
%the spectral changes of BO~Mic.
To investigate spectral changes 
we then derived  temperatures and volume
emission measures (EM=$\int n_{e}n_{H}dV$) of the individual plasma components
% with corresponding errors at 90\% confidence range
%for the different phases. Finally, we 
and calculated X-ray luminosities from the best fit models.
% Note that our error estimates
%of the OM observations do not take potential systematic effects into account.

\section{BO~Mic's activity spatially resolved}
\label{sec:Recon}
%\vspace*{-0.75\medskipamount}
\subsection{Doppler images}
\label{sec:DI}
\vspace*{-0.75\medskipamount}
%Reliable Doppler images (DI) form  the basis of any interpretation of 
%spatially resolved activity.
% and the
%construction of such images was one of the central goals of our \textsc{Vlt} spectroscopy.
The photospheric spots were reconstructed from the \textsc{Vlt} spectroscopy
using our Doppler imaging (DI) algorithm \textsc{Cldi} (WSW~2005)
on a grid resolving 2020 surface elements.
We adopted
an inclination angle of $70\degr\ $ for the stellar rotation axis, \mbox{$v\,\sin{i} = 134$~km/s}
as determined by WSW~2005 in agreement with \citet{Barnes05}, and a rotation period of
$0.380$~days (\citealt{Cutispoto97}).  
For DI, the spectra were added in pairs, resulting in 72 spectra with a typical S/N ratio 
of 400 at 6000~\AA.
The spectra cover 1.1 stellar rotations with a homogeneous 
phase sampling and a gap of 
%2.06 
two rotations between the observations of the two hemispheres.
The line profiles used for the DI 
were obtained by a spectrum deconvolution of the wavelength range 6390--6440~\AA\
%following the procedures described in detail in WSW~2005.
(see WSW~2005 for details).

\begin{figure}[h]
\center{

 \vspace*{1.5\medskipamount}

 \epsfig{file=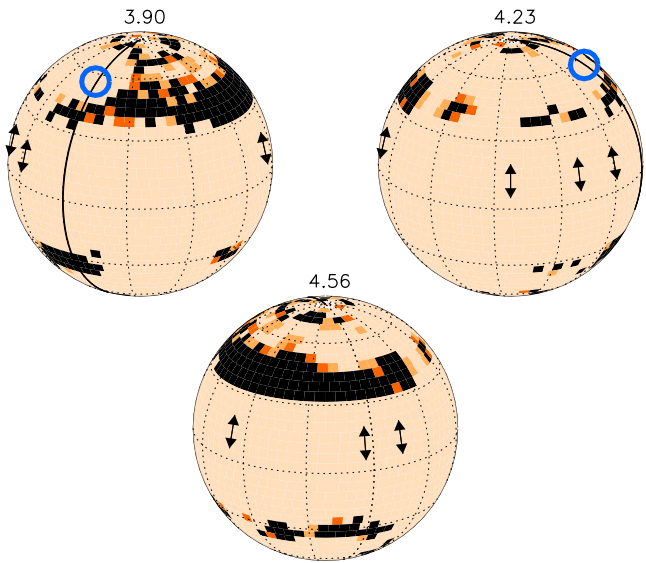, width=0.95\linewidth,clip=} 
}%center
\caption{
Atmospheric structures of BO~Mic  
%during a stellar rotation,
reconstructed by Doppler imaging and related methods.
All maps show the same features for the given rotation phases.
% above each map.
Photospheric spot coverage is rendered as black, dark and light orange areas,
representing 100\%, 67\% and 33\% spot coverage, respectively.
%Black squares on the map indicated photospheric regions with continuum flux 
%reduced by 50\%, corresponding to an average temperature difference of
Equatorial arrows mark approximate longitudes of prominences observed in Ca~K 
and \mbox{H$_\alpha$}. The blue circle indicates the
chromospheric emission region of the flare.
%It is shown shortly after its appearance and about 0.1~rotations before the flare
%disappears in X-rays.
Grid lines mark latitudes and longitudes in $30\degr$ steps.
%the zero meridian (passing below the observer at phase~0) is drawn thick.
        }
\label{fig:DI}
\end{figure}
%% created by spot_inspect_28a.pro, si_SpMic20067_surfs.par

\begin{figure}[h]
%\center{
\mbox{
% \hspace*{0.15cm}    %adjust for missing right y-axis annotation
 \epsfig{file=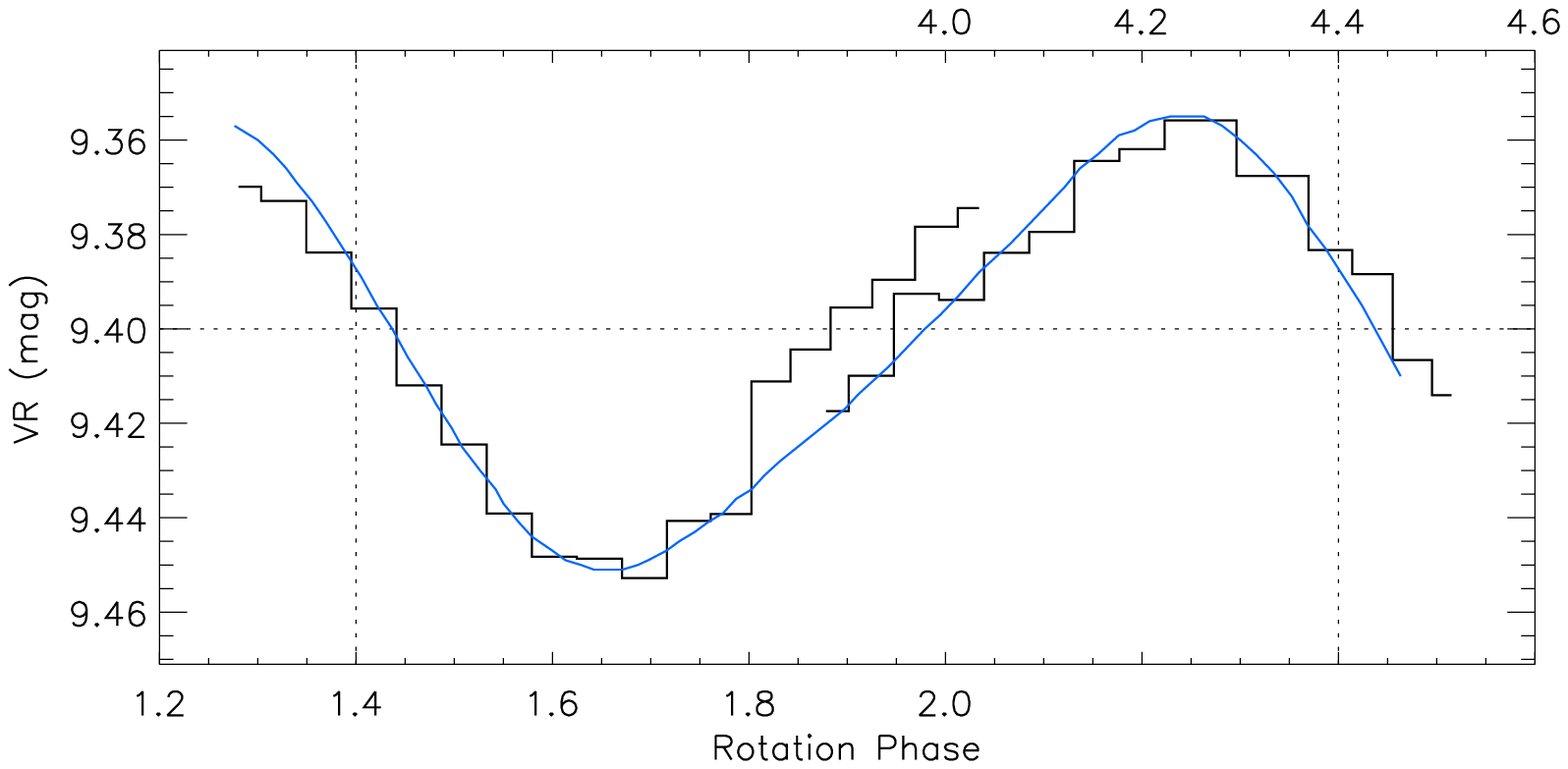, width=0.92\linewidth,clip=}  %0.92 
}
%}%center
\caption{
%BO Mic's emission from the soft X-ray to optical wavelengths as a function of rotation phase 
BO Mic's photospheric emission as measured by {\it XMM-Newton's} optical monitor (black)
and computed from our Doppler image 
%shown in Fig.~\ref{fig:DI} 
(blue).
The left and right sections of the black curve render the 
first (lower phase scale) and second (upper scale) 
observed semi-rotation, respectively.
                      The emission is integrated from  
                      6000 to 6500~\AA\,, i.e. centered
                      between Johnson-V and R.
                      The estimated 
                      error is $\pm 0.003$~mag.
                      %see Sec.~\ref{sec:Obs+Analysis} for details.
                      % although systematic errors related to the relative brightness of
                      % BO~Mic for OM observations
                      %for the first and second nights 
%        The symbols show a color index comparable to B-V with the symbols
%        representing the two stellar semi-rotations.  
%        The light curves are plotted with an offset 
%        to roughly match BO~Mic's apparent magnitude.
        %See text for discussion. 
        }
\label{fig:OptLcurves}
%\end{figure}

%\begin{figure}[t]
\center{
 \epsfig{file=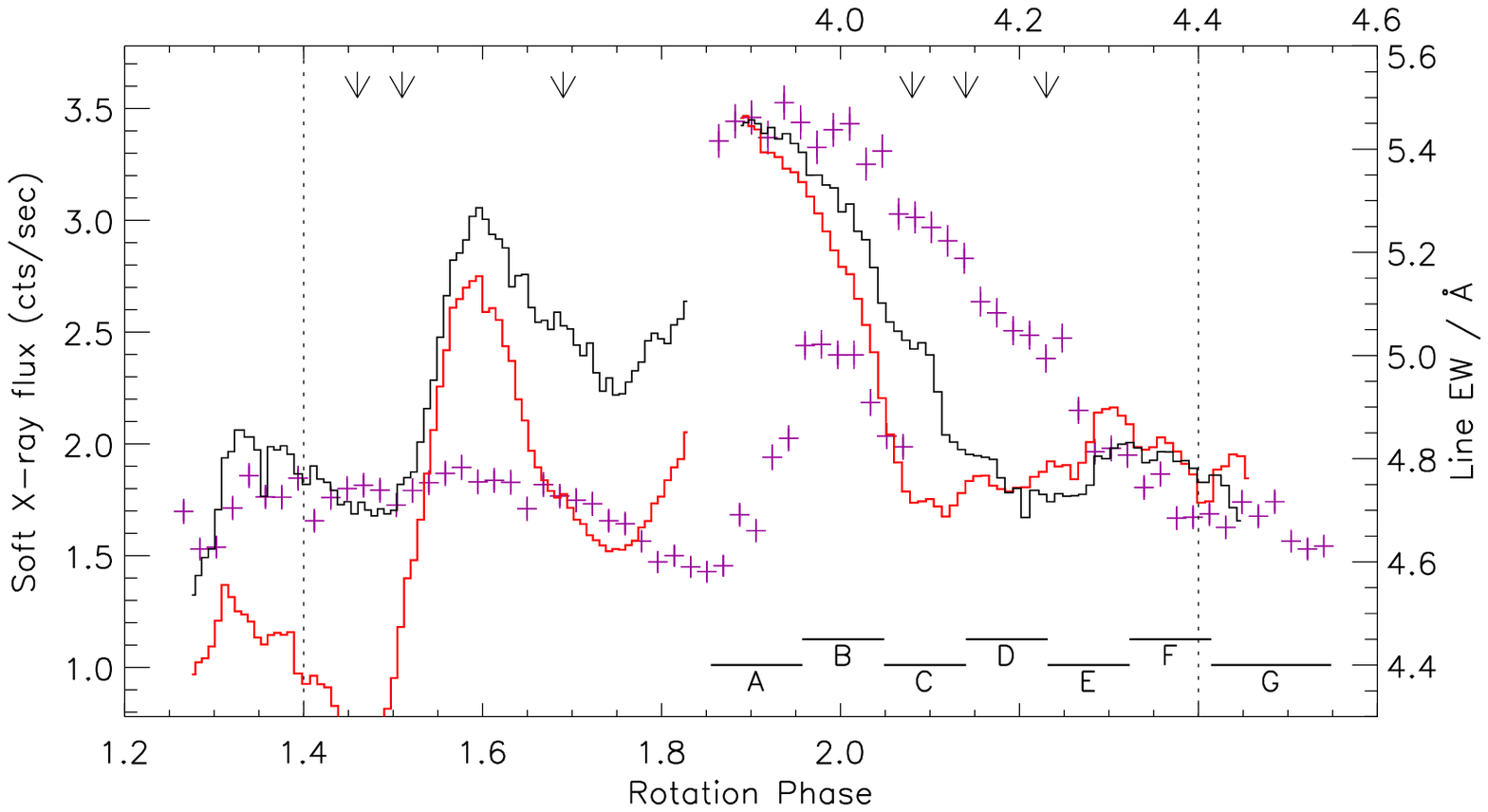, width=1.02\linewidth,clip=} %0.97 
}%center
\caption{
%BO Mic's emission from the soft X-ray to optical wavelengths as a function of rotation phase 
                      BO Mic's coronal and chromospheric emission.
                      Purple crosses show flux and \mbox{$1\sigma$-errors} of the soft X-ray emission ($0.2-5.0$~keV)
                      as registered by {\it XMM-Newton's} EPIC-pn detector.
                      The black and red curves give the 
                      Ca~K and \mbox{H$_\alpha$} equivalent width, respectively.
                      Arrows mark the phases of sub-observer passage for the prominences
                      shown in Figs.~\ref{fig:DI} and ~\ref{fig:CaK_profiles}.
     The horizontal lines ``A--G'' indicate bin intervals for the data points
     of Fig.~\ref{fig:Xray-T-EM}.
     Phase annotations are the same as in Fig.~\ref{fig:OptLcurves}. 
        }
\label{fig:XrayLcurves}
\end{figure}

In Fig.~\ref{fig:DI} we show a reconstruction of BO~Mic's active regions. 
%during our observations.  
The spots on the 
surface show a largely non-polar (similar to \citet{Barnes05}, although at lower latitude), 
azimuthally quite asymmetric
distribution, with spots both in the northern and southern hemispheres.
The large-scale reliability of our Doppler image,
at least in longitudinal spot distribution,
is illustrated in Fig.~\ref{fig:OptLcurves},
where we compare the observed light curve with that computed from our DI in the 6000--6500~\AA\, 
range. The light curve supports the reality of the
southern spots found in the DI. 
While their appearance
in the DI depends on the adopted spot brightness,
%($50\%$ of the undisturbed photosphere for the 
%DI of Fig.~\ref{fig:DI}), 
the reconstructions including southern spots
yield a significantly better fit to the OM-light curve.
%In agreement with other DI of BO Mic (\citealt{Barnes05}) the polar region is
%only weakly spotted on our image.

%% created by 
Concerning the non-periodic components of the OM-light curve, we note
that the peaks at phases 1.25 and 4.25 differ in brightness.
This reflects a change in the spot pattern during the three rotations between them,
% This is
as supported by changes in the corresponding photospheric line profiles.
% and X-ray emission. 
%in the phase overlap region.
%Unfortunately, we do not know what causes the abrupt brightness increase at phase 1.8.
% during the first night.
% between the two observed semi-rotations.
\subsection{The chromospheric and coronal perspective}
\vspace*{-0.75\medskipamount}
In Fig. \ref{fig:XrayLcurves} we plot the temporal evolution of the chromospheric and
coronal activity in terms of Ca K and H$_{\alpha}$ equivalent width 
and soft X-ray emission in the $0.2-5$~keV energy band (crosses).  
Clearly, neither varies smoothly 
%at the transition 
between the two nights.
%, and in X-rays the emission level (at the same phase) approximately doubled in the second night.
The X-ray light curve 
%in Fig. \ref{fig:XrayLcurves} 
shows flares during both nights;
%light curve of the whole 
the second night
seems to consist of one 
%more or less 
coherent event 
%lasting throughout most of the observations,
with several bumps in the decaying X-ray light curve, 
suggesting reheating events. 

\begin{figure}[h]
\center{
 %\vspace*{1.5\medskipamount}
% \epsfig{file=figs/JR/speedyfl2.epsi, width=0.95\linewidth,clip=}
% \epsfig{file=figs/JR/speedyfl2e.ps, width=0.95\linewidth,clip=} 
 %\epsfig{file=speedyfl2.ps, width=0.95\linewidth,clip=} 
 \epsfig{file=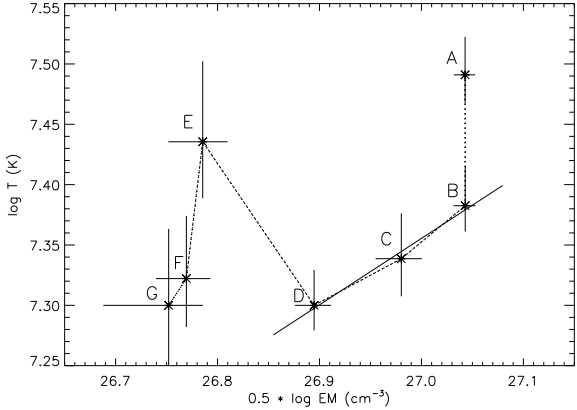, width=0.90\linewidth,clip=}  %ref2 shortened: 0.95-0.90
}%center
\vspace*{-1.0\medskipamount}
\caption{
Evolution of emission measure and temperature for the hot plasma component
during the flaring, with $1\sigma$ errors as determind by the 
X-ray spectral fit. 
EM$^{0.5}$ is used as a proxy for the plasma density.
The solid line shows a linear fit to points B--D, corresponding to the decay
of the first flare event. Further energy release is evidenced by the
deviation of points E--G from the decay path.
%Phase ranges of the 
The time bins of points ``A--G'' are given in Fig.~\ref{fig:XrayLcurves}.
        }
\label{fig:Xray-T-EM}
\end{figure}

To verify this interpretation we investigated the
evolution of the X-ray emitting plasma
in terms of emission measure and temperature.  We divided the X-ray data into seven segments 
(``A-G'' in Fig.~\ref{fig:XrayLcurves}) and performed a 
spectral analysis 
%as described in 
(cf. Sect.~\ref{sec:Obs+Analysis}).
The results are shown in Fig.~\ref{fig:Xray-T-EM}:
the temperature decreases with decreasing
emission measure during the intervals A-D, 
which is consistent with the decay phase of a flare.
During interval E the temperature rises, 
indicating additional energy input. 
%The observed plasma temperatures of 20\,--\,30~MK are typical for moderate flares on active stars.
The temperatures of 20\,--\,30~MK are typical for moderate flares.

%We fitted the decay of the first event, and used the derived slope and the method described
%in \cite{rea04} for a single loop to determine an upper limit for the loop half length.

%fluxes approximately doubling.
%increasing by about a factor of two, 
%More pronounced 
%in the flare 
%during the second night, it coincides with the event discussed 
%n the previous section.
%Both flares are observed at similar rotational phases 
%(around $\phi \approx .0$).
% suggesting that a related stellar surface region is responsible for the enhanced X-ray 
%activity in both nights. 
%Spectral analysis shows that the enhanced flux level at these phases cannot be described 
%by a single flare, but rather has to be attributed to on overlay of several consecutive flaring events. 
Outside the flaring phases the X-ray light curve is relatively flat
and apparently dominated by a stable quasi-quiescent level;
it varies only by 20\% peak to peak.
%superimposed on 
As visible in Fig.~\ref{fig:XrayLcurves}, this modulation has a local maximum centered
on phase $\phi \approx 1.55$, so it may be correlated with the
photospheric brightness minimum at $\phi \approx 1.65$ (Fig.~\ref{fig:OptLcurves}).
However, based on one observed rotation, and in the presence of flaring, this
%conclusion 
remains speculative.

\subsection{Localizing prominences and the flare}
\label{sec:Prominences}
\vspace*{-0.5\medskipamount}

\begin{figure}[t!]
\center{
 \vspace*{-0.5\medskipamount}
  \mbox{
  \hspace*{-0.3cm}    
   \epsfig{file=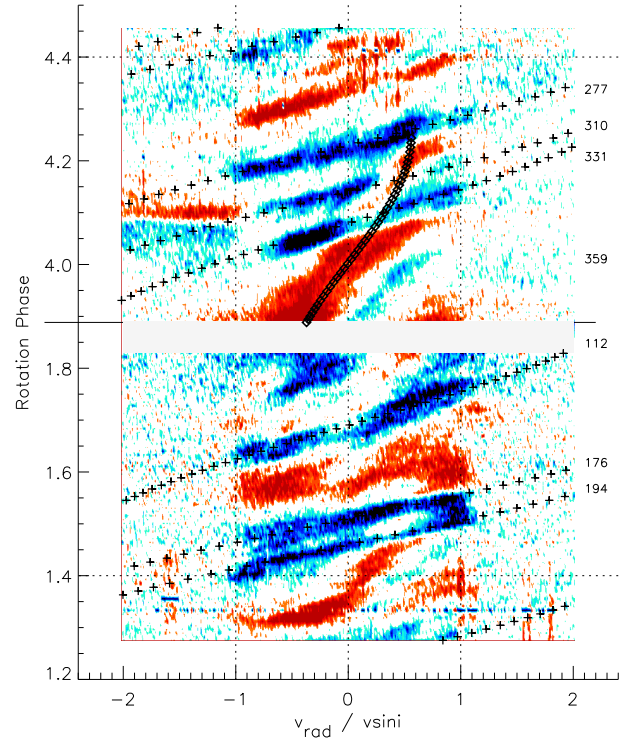, width=0.9\linewidth,clip=} 
         %bb= 70 280 460 730} 
  }
}%center
\vspace*{-1.0\medskipamount}
\caption{
Evolution of the Ca~K emission:
%\textit{Upper panel:} 
%Line profiles observed before and during the flare, phases are given to the
%left of the diagram.
%The profiles are approximately normalized to the continuum surrounding the Ca~K line.
%Diamonds mark the peak positions of the flare-related emission,
%as modelled by the sine-fit in the panel below.
%\textit{Lower panel:} 
Ca~K line core profiles, processed by unsharp masking
are rendered as a function of wavelength (units of $v\,\sin{i}$)
and rotation phase.
%Wavelengths have been transformed to units of
%\mbox{$v\,\sin{i} = 134$~km/s}.  
Absorption dips in the profile appear blue while
emission bumps are shown red. 
Plusses indicate sine-fits according to Eq.~\ref{eq:surf_feature_pos}. They are  used to
localize absorbing prominences, %passing the stellar disk, 
their longitudes are given to the right.   %right of each fit.
The fit to the flare-induced chromospheric emission is marked by diamonds.
See text for discussion.
        }
\label{fig:CaK_profiles}
\end{figure}

In Fig.~\ref{fig:CaK_profiles} we provide a detailed view of the evolution of the Ca~K emission
line profile during our observations.  Specifically, 
%Figure~\ref{fig:CaK_profiles} shows
we show the time series of line profiles processed by an unsharp-masking filter that
enhances fast changes in the core profile.
The unsharp masking is performed by dividing the profile time series by a temporally
smoothed copy of itself.  For Fig.~\ref{fig:CaK_profiles}, wavelengths were transformed to 
velocity units, absorption dips in the profile are shown in blue while
emission features are shown red.  Several absorption features clearly
move through the Ca~K line profiles.
% in Figure~\ref{fig:CaK_profiles}.  
In order to localize the corresponding prominences in longitude
we fit pronounced deformations migrating redward through the core profiles
with sine functions depending on rotation phase $\phi$ through
\begin{equation}
\label{eq:surf_feature_pos}
v_\mathsf{rad}(\phi) = v\,\sin{i} \cdot r/R_{\ast} \; cos{(\pi/2 - \theta)} \, sin{(\varphi - 2\pi \phi)} 
\end{equation}
which describes the radial velocity of an atmospheric feature
located at the radius~$r$ (divided by the stellar radius $R_{\ast}$), the
latitude~$\theta$ and longitude~$\varphi$. 
Essentially, the phase of $v_\mathsf{rad}=0$ determines the longitude
while the slope $\frac{dv}{d\varphi}$ determines the height of a prominence 
when $\theta$ is adopted as the subobserver latitude.
In this way 
we obtain the approximate longitudes and heights of six prominences 
marked in Fig.~\ref{fig:CaK_profiles}.
These prominence parameters agree with those obtained by the 
same procedure from the \mbox{H$_\alpha$} core profiles.
Their radii range between 2.5 and 3.5~$R_{\ast}$ with an uncertainty
of \mbox{$\pm 0.5\,R_{\ast}$},
i.e. they are about 2~$R_{\ast}$ above the surface.
In Figs.~\ref{fig:DI} and~\ref{fig:XrayLcurves} the 
approximate longitudes and phases of the structures causing 
absorption transients
in the Hydrogen Balmer and 
\mbox{Ca\,H} and K lines are marked by arrows. 
Such fast-moving, often periodic line profile deformations 
%in  \mbox{H$_\alpha$} 
can be attributed to ``prominences'', i.e.
clouds of relatively cool material magnetically kept in co-rotation high
above the stellar surface, and
%which have previously been 
already
observed on BO~Mic (\citealt{Collier03}; \citealt{Dunstone06}).
%\subsection{Localizing a flare-induced plage}
%\label{sec:plage-loc}
%\vspace*{-0.5\medskipamount}
Given their rapid movement through the line profile,
these prominences cannot be located close to the stellar surface.
Only one pronounced deformation of the Ca~K line core moves through the
profile compatible with an  {\bf emission} feature close to the stellar surface.
Shown in Fig.~\ref{fig:CaK_profiles}, it passes the profile center at phase 
%\mbox{$\phi \approx 4.0$}.
\mbox{$\phi=4.0\pm0.02$}.
Using Eq.~\ref{eq:surf_feature_pos} to trace the emission peak
through the line profile, we locate its emitting
region on the surface at a latitude of \mbox{$\theta = 56\pm10\degr\ $} 
and a longitude of
\mbox{$\varphi = 359\pm 10\degr$}, marked by a circle
in Fig.~\ref{fig:DI}.
%The corresponding sine-fit is shown in Fig.~\ref{fig:CaK_profiles}.
%Additionally, the peak positions are marked in the line profiles
%of the upper panel of the same figure.

%  Returning to the flare of the second night,

It should be noted that this ``flare plage'' could also be located at the same latitude 
south of the equator, since for each rotation phase the radial velocity
is the same there  as for the northern location.
The same effect causes a north-south-ambiguity
of features found in all Doppler images (e.g. \citealt{Pisk90}).
However, in that case, the weak emission rise in \mbox{H$_\alpha$} and Ca~K
at phase $\phi \approx 4.3$ (see Fig.~\ref{fig:XrayLcurves}) 
could not originate from the plage location
because south of the equator this latitude is no longer
visible.  Thus, this feature is very likely located in the north.
%Since the observed Ca K and H$_{\alpha}$ emissions seem to follow (or lead) 
%the X-ray light curve
Since the enhanced Ca K and H$_{\alpha}$ emissions clearly 
coincide with the X-ray flare 
%it is thus natural associate these also with the X-ray flare.
%
%Since it is precisely coincident with the X-ray flare  
%
%the feature is in emission, it is present at the very beginning of the second night,
%when the X-ray flare is ongoing, 
we are confident that 
%the association between Ca K emission and X-ray flare is real, and that 
the location marked in
Fig.~\ref{fig:DI} is indeed the chromospheric footpoint of the X-ray flare.

\section{The flare: evolution and energetics}
\label{sec:FlareIntro}
\vspace*{-0.75\medskipamount}
Our contemporaneous optical and X-ray data allow us to further characterize the
flaring plasma.  Unfortunately, our observations did not cover the impulsive rise
of the chromospheric and coronal flare emission and we see no flare signatures 
in the photometric lightcurves.
However, a short increase prior to the peak of the X-ray emission is visible in Fig.~\ref{fig:XrayLcurves}, 
suggesting that our observations
cover the very end of the rise phase of the flare and all of its decay.
The decay to the ``quiescent'' level takes about 0.45 and 0.29 rotation phases or 4.1 and 2.6~hours for the soft X-ray and Ca~K
emission, respectively.
The \mbox{H$_\alpha$} emission decays even faster, however, it is 
strongly influenced by the prominences.

Assuming that the flare occurs in a single plasma loop and using the method described in \cite{rea04},
we estimate the flare loop size from the EM-T diagram shown in Fig.~\ref{fig:Xray-T-EM}.
Fitting the decay of the first heating event, the derived slope
yields an upper limit of \mbox{$l \la 0.4\,R_{\ast}$} for the loop half length.
%--ok ref2: insert
This upper limit even remains valid if the flare occurred in an arcade of 
several loops, because in this case each individual loop would have to be smaller.
While considerably larger than solar coronal loops,
this is much closer to the star 
than the prominences located at $h\approx2\,R_{*}$
(cf. Sect.~\ref{sec:Prominences}). Reheating occurs between rotation phases
4.2 and 4.3, indicating its connection to the sudden slight increase
in X-rays seen in Fig.~\ref{fig:XrayLcurves} at phase
4.25. This 
%X-ray increase 
is possibly related to the increase in
\mbox{H$_\alpha$} and Ca~K emission observed about 0.05 rotations
\mbox{($\approx 30$~min.)} later.
% At any rate,
The total emitted energy of the flare observed in soft X-rays, \mbox{$\approx 1.3\cdot10^{34}$~erg} in the 0.2\,--\,10~keV band,
appears to originate predominantly, but not exclusively, from a single energy release.
The peak luminosity of the flare in soft X-rays of 
\mbox{$\approx 1.4\cdot10^{30}$~erg/s} ($6\times10^{29}$~erg/s in the 1--8\,\AA~GOES-band) 
%ref2 shortened: used by GOES) 
is considerably larger
than the maximum power
%ref2 shortened: radiated 
in Ca~K and \mbox{H$_\alpha$}
of \mbox{$\approx 2.0\cdot10^{28}$~erg/s} and \mbox{$\approx 1.0\cdot10^{29}$~erg/s},
respectively.
This energetic mismatch of coronal and chromospheric emission
would be unusual for a solar flare where these emissions --
with considerable variations -- tend to be comparable
(\citealt{Foukal04}, \citealt{Johns-K97}, \citealt{Emslie05}).
%We finally note that in the broadband light curve
%no flare signatures can be detected, however, the actual flare onset has not been 
%covered by our observations.

\section{Summary and discussion}
\label{sec:Disc}
\vspace*{-0.75\medskipamount}

We observed the highly active K-star BO~Mic during one rotation
with completely simultaneous coverage from soft X-ray to optical wavelengths.
The observations were carried out in two blocks separated by two stellar rotations
with all observed indicators of photospheric, chromospheric and coronal activity 
%of BO~Mic 
showing pronounced variations.  
%The photospheric light curve is
%roughly sinusoidally modulated with the previously known rotation period. 

The Doppler image (DI)
%, derived from the spectra, 
shows only a few spots near the visible pole.
Most spots are asymmetrically distributed at mid-latitudes, 
%with a high degree of azimuthal asymmetry, 
leading to 
a roughly sinusoidally modulated photospheric light curve with 
a peak-to-peak amplitude of $\approx$ 0.1 mag.
There are additional small non-periodic components in the optical light curve; 
%during the observed three rotations;
in agreement with results of previous observations of BO~Mic 
(WSW~2005), this indicates some reconfigurations of spots on rather short
 timescales of a few rotations.

The chromospheric light curves (\mbox{H$_\alpha$} and Ca II) and coronal light curve
 (soft X-ray emission) are only weakly correlated with the 
photospheric modulation.  The predominant variations of chromospheric emission
are absorption transients in the \mbox{H$_\alpha$} and Ca~K line cores,
which are caused by prominences about two stellar radii above the surface;
their distribution is qualitatively similar to those analyzed for
BO~Mic by \cite{Dunstone06}. The longitudes of the prominences show no obvious connection to
the photospheric spots.
Whether the ``quiescent'' chromospheric emission is more intensive
on the photospherically darker hemisphere, as suggested by our
2002 observations (\citealt{Wolter05a}), cannot be reliably determined due
to the strong prominence absorptions found here.  Also, there is no clear evidence for X-ray absorption
by these prominences, setting a strict upper limit to their column density.

The most conspicuous event during our observations was a flare lasting about 4 hours; the flare
is not registered in our broad band photometry, but shows up clearly in the \mbox{H$_\alpha$}, Ca II and
X-ray light curves. By tracing the radial velocity of its chromospheric
emission in the Ca~K line we localize the flare site in the stellar atmosphere.  Somewhat surprisingly,
the flare does not appear to be connected to the main concentration of activity in terms of dark spots;
rather, it is located at the fringes of this region.
As discussed in Sect.~\ref{sec:Prominences}, this ``flare plage'' 
(which could also be a post-flare loop)
could also be situated
at the same latitude south of the equator, where the DI is only poorly 
defined due to poor visibility close to the stellar limb.
At any rate, the ``flare plage'' is not associated with any distinguished feature of the
photospheric spot pattern. Either such a spot feature is largely hidden
by the southern limb or, more likely, it does not exist. 

We determine the overall energetics and thermal evolution of the flare 
by simultaneously studying the soft X-ray emission.  With
a peak soft X-ray power of \mbox{$\approx 10^{30}$~erg/s}, this flare is
a hundred times more energetic than a large solar flare in terms of peak flux
and total radiated energy (\citealt{Golub97}),
however, it is still at least two orders of magnitude weaker than
the largest flares observed on BO~Mic.
%To our knowledge there are no observations of flares
%on BO~Mic using optical photometry; 
Assuming \mbox{$\approx 10^{30}$~erg/s} as an upper limit to the
power radiated in the optical continuum, 
%we compute a corresponding fraction of
this would correspond to 
$\sim 10^{-3}$ of BO~Mics total luminosity (WSW~2005),
i.e. below
the sensitivity of our photometry and consistent with the 
%flare's 
photometric non-detection.

In addition to the large flare, BO Mic's soft X-ray emission exhibits
another, smaller flaring event and possibly a slow modulation with \mbox{$\approx 20\%$}
peak to peak amplitude.
The slow modulation may be correlated with the observed photospheric spot
distribution; however, due to the flares and only one observed 
rotation, this remains inconclusive.  Also, no optical spectra are available
for the first flare, therefore it could not be localized.
However, both X-ray flares appeared at
similar rotation phases ($\phi \approx 2.0$ and~$4.0$), 
i.e., the first flare was situated
on the same hemisphere as the second or sufficiently
close to one of the poles.
Qualitatively, our X-ray lightcurve is similar to that observed
by \cite{Hussain07} for the ultrafast rotating K-star AB~Dor.
Also, our value for the coronal flare loop height
of \mbox{$\la 0.4\,R_{\ast}$} is compatible with their height estimate of coronal structures,
%on AB~Dor, 
once more confirming the similarity of AB~Dor and BO~Mic.
%Interestingly, \cite{Maggio00} find similar plasma loop heights for 
%two much more energetic flares on AB~Dor.
%As \cite{Maggio00} we find no correlation between the prominence locations
%and the flaring region.

To our knowledge, the flare on BO~Mic is the first X-ray flare that could be localized on
a Doppler image.  Interestingly, the flare occurs on a rather inconspicuous portion
of the atmosphere as judged from the image.  Extended observations of flaring events
on a given star could reveal the spatial distribution of flare emerging
sites. Physically, one expects flaring events to be associated with magnetic null lines.  Since 
our UVES spectra are non-polarimetric, we cannot assess the location of magnetic
null lines on BO Mic's surface.  Clearly, a Zeeman Doppler image would be helpful to prove
the association of stellar flare sites with magnetic null lines as expected from the solar analogy.

%This modulation is very closely modelled by the lightcurve computed from
%our Doppler images confirming the large-scale reliability of the images.

% ok-checked 24.9.  ckeck prominence -- soft X-ray correlation again

% SCRATCHPAD: What do we learn from the induced flare location? ... associated with the large spot group ...
%possibly with the pronounced and somewhat concentrated spot group reconstructed in our
%DI on the southern hemisphere ... reality not certain, but supported to some degree by
%the optical lightcurve. ... rather high latitudes, i.e. close to the pole ...
%possible flare height (X-ray diagnostics) would allow sufficient X-ray emission
%visibility even on southern hemisphere?
%  check this for chromospheric emission
%plage localization and chromospheric vs. rotational velocities
%   blue wing - red wing - blue wing absorption feature ... companion ? ... Barnes 2005 Fig. 2
%DI characteristics, north-south (flare visibility --> north (?)), spot contrast:
%more detailed DI-study
%   no obvious correlation of spot distribution and chromospheric emission in contrast
%   to 2002 observations (WS205 Fig. 4)
%However, ... suggested re-appearing emission regions ... compare to Hussain 2007 Fig.3 

\begin{acknowledgements}
     % U.W. acknowledges financial support from
     % \emph{Deut\-sche For\-schungs\-ge\-mein\-schaft}, \mbox{DFG - for the different phases
     % SCHM 1032/21-1}.
J.R. and U.W. acknowledge DLR support \, (50OR0105).
\end{acknowledgements}


\begin{thebibliography}{}
%\bibitem[X, 2006]{X06}
\bibitem[Arnaud, 1996]{xspec96}
%{Arnaud}, K.~A. 1996, in ASP Conf. Ser. 101: Astronomical Data Analysis Software and Systems V, ed. G.~H. {Jacoby} \& J.~{Barnes}, 17
Arnaud K.~A. 1996, in ASP Conf. Ser. 101: Astronomical Data Analysis Software and Systems V, ed. G.~H. {Jacoby} \& J.~{Barnes}, 17
\bibitem[Barnes, 2005]{Barnes05} Barnes J.R., 2005, MNRAS 364, 137
\bibitem[Collier Cameron et al., 2003]{Collier03} Collier Cameron A. C., Jardine M.,
%Wood K., Donati J.-F, 
et al.,
2003, EAS Publ. Ser. 9, 217 
\bibitem[Cutispoto et al., 1997]{Cutispoto97}  %%multicolour photometry and period of SpMic
  Cutispoto G., 
  %K\"urster M., Pagano I., Rodono M., 
  et al.,
  1997,
  Information Bulletin on Variable Stars, 4419, 1
\bibitem[Dunstone et al., 2006]{Dunstone06} Dunstone N.J., Barnes J.R. 
%Collier Cameron A. C., Jardine M.,
et al., 2006, MNRAS 365, 530

\bibitem[Emslie et al., 2005]{Emslie05} Emslie A.G., 
         %Dennis B.R., Holman G.D., Hudson H.S., 
         et al.,
         2004, Journal of Geophysical Research, 110, A11103

\bibitem[Johns-Krull et al., 1997]{Johns-K97} Johns-Krull C.M., Hawley S., Basri G., 
                         Valenti J.A., 1997, ApJS, 112, 221  

\bibitem[Foukal, 2004]{Foukal04} Foukal P., Solar Astrophysics, 2004, Wiley, New York

\bibitem[Golub \& Pasachoff, 1997]{Golub97} Golub L., Pasachoff J.M., 
                 The solar corona, 1997, 
                 Cambridge University Press  %, Cambridge

\bibitem[Hauschildt et al., 1999]{Haus99} Hauschildt P., Allard F., Baron E.,
                         1999, ApJ, 512, 377 

\bibitem[Hussain et al., 2007]{Hussain07} Hussain G. A. J., Jardine M., Donati J.-F.,
                         %Brickhouse N. S., Dunstone N., et al., 
                         et al.,
                         2007, MNRAS, 377, 1488

\bibitem[K\"urster, 1995]{Kurster95} K\"urster M., 1995, Flares and Flashes, IAU Coll. 151, 
  p. 423

\bibitem[{{Loiseau} {et~al.}(2006){Loiseau}, {Ehle}, {Pollock}, {Talavera}, {Gabriel}, \& {Chen}}]{sas}
{Loiseau}, N., {Ehle}, M., {Pollock}, A., {et~al.} 2006

\bibitem[Maggio et al., 2000]{Maggio00} Maggio A., Pallavicini R., Reale F., Tagliaferri G.,
             2000, A\&A, 356, 627

\bibitem[{{Reale} {et~al.}(2004){Reale}, {G{\"u}del}, {Peres}, \& {Audard}}]{rea04}
{Reale}, F., {G{\"u}del}, M., {Peres}, G., \& {Audard}, M. 2004, \aap, 416, 733

\bibitem[Piskunov et al., 1990]{Pisk90} Piskunov N., Tuominen I., Vilhu O., 1990, A\&A, 230, 363    %DI reconstruction 

\bibitem[Piskunov \& Valenti, 2002]{Piskunov02} Piskunov N., Valenti J.A., 2002, A\&A, 385, 1095    %Reduce

\bibitem[Schrijver \& Zwaan, 2000]{Schrijver00} Schrijver C.J.,
  Zwaan C., 2000, Solar and Stellar Magnetic Activity, 
  %Cambridge University Press
  CUP

\bibitem[{{Smith} {et~al.}(2001){Smith}, {Brickhouse}, {Liedahl}, \& {Raymond}}]{apec}
{Smith}, R.~K., {Brickhouse}, N.~S., et al.,
%{Liedahl}, D.~A., \& {Raymond}, J.~C. 
2001, \apjl, 556, L91

\bibitem[Wolter \& Schmitt, 2005]{Wolter05a}
  Wolter U., Schmitt J.H.M.M., 2005, A\&A, 435, L21
%\bibitem[WSW, 2005]{Wolter05}
\bibitem[Wolter et al., 2005]{Wolter05}            %why double comma? - it works with citealt
  Wolter U., Schmitt J.H.M.M., van Wyk F., 2005, A\&A, 435, 261

\end{thebibliography}
\end{document}